# Drop on demand in a microfluidic chip

Jie Xu and Daniel Attinger

*Laboratory for Microscale Transport Phenomena, Department of Mechanical Engineering, Columbia University, New York, NY 10027, USA*

In this work, we introduce the novel technique of *in-chip drop on demand*, which consists in dispensing picoliter to nanoliter drops on demand directly in the liquid-filled channels of a polymer microfluidic chip, at frequencies up to 2.5 kHz and with precise volume control. The technique involves a PDMS chip with one or several microliter-size chambers driven by piezoelectric actuators. Individual aqueous microdrops are dispensed from the chamber to a main transport channel filled with an immiscible fluid, in a process analogous to atmospheric drop on demand dispensing. In this article, the drop formation process is characterized with respect to critical dispense parameters such as the shape and duration of the driving pulse, and the size of both the fluid chamber and the nozzle. Several features of the *in-chip drop on demand* technique with direct relevance to lab on a chip applications are presented and discussed, such as the precise control of the dispensed volume, the ability to merge drops of different reagents and the ability to move a drop from the shooting area of one nozzle to another for multi-step reactions. The possibility to drive the microfluidic chip with inexpensive audio electronics instead of research-grade equipment is also examined and verified. Finally, we show that the same piezoelectric technique can be used to generate a single gas bubble on demand in a microfluidic chip.

## *Introduction*

The concept of lab on a chip, where a tiny fluid microprocessor performs complex analysis and synthesis tasks relevant to chemistry or biology, has been a subject of academic interest since the early 1990's [1, 2], and there are encouraging signs that industrial lab on a chip applications are growing, exemplified by microfluidic devices for DNA sequencing [3]. The operations needed for biological or chemical analysis are reagent dosing, transport, mixing, splitting, flushing, filtering, analysis, detection and monitoring [4], each operation requiring a precise controll in space and time [5]. The microfluidics components needed do perform these operations are buffers, channels, valves, mixers, microheaters and sensors [4-6]. Several achievements, mostly in academia, have demonstrated that shrinking a chemical or biological laboratory into a microchip could have significant benefits such as increased sensitivity, fast response time, low reagent and sample consumptions, as reviewed in [2, 6-8]. The ability to dispense and control small liquid volumes in the microchannels is critical for the lab on a chip technology, and several techniques address, at least partially, this issue. For instance, electrokinetic pinching [9] or the volumetric change induced by a piezoelectric actuator [10] have been used to inject individual liquid plugs in a miscible liquid, followed by successful electrochemical analysis [9]. Also, the segmented flow technique, defined as the transport of two or more immiscible phases in the form of successive plugs in a microchannel, has been developed: it uses syringe pumps feeding two branches of a T-connection [5, 11, 12] or two concentric channels [13, 14], as reviewed in [5]. In the T-connection process, drops of the so-called *dispersed* phase are produced in the other liquid phase, the *continuous* phase, as a result of shear forces and interfacial tension at the fluid-fluid interface [15, 16]. The process using two concentric tubes, called *flow focusing*, initiated by Ganan-Calveo [17], injects the disperse phase from the smaller tube into the continuous phase that flows out of the larger tube, and breakup occurs in an analogous manner as the breakup of an atmospheric jet in atmosphere, which is due to the Plateau-Rayleigh-Savart instability [18-20]. The segmented flow technique decreases mixing times by several orders of magnitude [5, 15], while ensuring an extremely low diffusion between the two immiscible phases [5, 21]. These two unique features are extremely attractive for studying and controlling the timing of chemical reactions [21]. Other tasks successfully demonstrated with the segmented flow technique are the splitting of one drop into two half-size drops using a T-connection [22] and the electrical sorting of drops in the microchannels [23]. A couple of microfluidics systems have been produced to dose an individual quantity of liquid in the nL-µL range [24-31], to be used outside the chip, in a manner analogous to ink-jet printing. Recently, high-performance syringe pumps (microinjectors) [32, 33] or a high-voltage pulse have been used to induce the formation a single drop of liquid at a microfluidic T-junction [34, 35], in a dripping-like process. While controlling the drop volume [32, 34] and generation timing [35] precisely was difficult, the techniques could be used to encapsulate single cells [32]. In parallel to techniques manipulating droplets in chips, digital microfluidics, also called electrowetting on dielectric, has been developed, which now successfully generates, merges, transports, splits and heats drops with typical volumes of 50 nL and above, at the surface of a microfluidics chip [4].

The segmented flow and digital microfluidics techniques have the advantage over electrokinetic pinching that the diffusion between the minute amount of liquid dispensed and the carrying fluid is minimal. Both methods are suited for performing complex, multistep analysis or synthesis [4, 21]. Some drawbacks of digital microfluidics in its current state are presented by Fair in [4], such as the need to use conductive liquids, the relatively large volume of the drops (50 nL and above), and the inability to perform electrokinetic separation useful for chemical characterization. The drawbacks of segmented flow techniques are that a setup might need a relatively long accommodation time before generating the train of particles (drops or bubbles) in a stable manner [36], the difficulty to reliably generate a single particle rather than a train of particles [32-35], and the fact that segmented flow techniques are not as flexible as digital microfluidics for processing multistep reactions [37]. There is therefore a need for a technique that can generate a single particle, on demand, in an immiscible fluid. This need is acknowledged by George Whitesides, who fathered soft microfluidics: "There is a particular bit of the puzzle that needs to be added, which will not be hard to do but it has not been done yet—that is, bubble on demand [38]". In this paper, we present a technique to dispense a single drop or bubble on demand in a microfluidic chip. Advantages of this drop and bubble on demand technique over continuous segmented flow techniques can be suggested by looking at the history of atmospheric ink-jet printing [39] and other drop on demand dispense techniques [40-43], where continuous ink-jet printers have been outdated in the 1970s by the invention of drop on demand printers, which eliminate the need for sorting undesired drops.

In section 1 of this article, we describe the novel technique to dispense minute, individual amounts of one fluid into another immiscible fluid by ways of piezoelectric actuation, a technique tentatively called *in-chip drop on demand*. In section 2, we characterize the dispensing technique. Section 3 describes five features of this technique with direct relevance to the lab on a chip community.

## *1. Design, fabrication and setup description*

The typical design of a microfluidic chip used in our study is shown in Figure 1a: it involves one or several µL-volume reagent chambers such as *B* connected via a 25-100 µm nozzle to a main channel *A*. Using syringes, the main channel is filled with a fluid such as hexadecane while each chamber *B* can be filled with a different aqueous reagent. Since the water-hexadecane system is immiscible, a stable meniscus forms at the nozzle opening. The height of the channels is in the 50-100 µm range. The chip is sealed with a thin membrane. A piezoelectric actuator is placed on top of each chamber, to modify the chamber volume and release an aqueous drop, on demand, in the main channel. This drop generation process is shown in Figure 1b and the associated movie [44]. Once in the main channel, the drop can be transported using viscous drag towards the shooting area of another nozzle, where a drop of another reagent can be dispensed and mixed to the initial drop. This way, a sequence of reactions can be performed by bringing the original drop in front of several chambers shooting prescribed reagents. The internal flow associated with the motion of drops in a channel [10] enhance mixing and diffusion. The needed reagent volume (or *dead volume*) is relatively small, for instance a chamber of 20 mm×5 mm×50 µm fed by a tube of 300 µm diameter and length L=1cm represents a dead volume of 8 microliter, which corresponds to enough ammunition to shoot 10,000 80-pL drops.

The microfluidic chips are fabricated in the clean room of Columbia University using soft lithography [45]. First, a 10 µm thin base layer of *SU-8* resin (MicroChem) is spun and cured on a silicon wafer. On top of that layer, a 50-100 µm layer of *SU-8 2050* is cured with patterns transferred from a mask (CAD/Art Services Inc.). This base layer method presented in Carlier et al. [46] improves adhesion of *SU-8* to the wafer. The chip is then manufactured from the master using PDMS Sylgard 184 Kit (Dow Corning). The channels are sealed by a thin 180 µm membrane made from spin-coated PDMS. The piezoelectric actuators are commercially available bimorph actuators made of two PZT layers bonded on a thin brass layer, with a total thickness *T* 0.51 mm, with lengths and widths slightly smaller than the chamber dimensions as shown in Figure 1a and as given in Table 1. One actuator is then taped on top of each chamber, using a 90 µm layer of double-sided tape.

The experimental setup shown in Figure 2 involves three subsystems: the microfluidic system, the actuation system and the sensing system. The microfluidic system involves the microfluidic chip described above (letter *a* in Figure 2). Syringes fill the main channel with hexadecane and control the subsequent injection of aqueous plugs in the dispensing chamber. The actuation system uses a 20MHz function generator (*letter d,* Agilent, 33120A) coupled to a 1MHz 17W amplifier (*letter e,* Krohn-Hite, 7600M), which generates high-voltage driving pulses for the actuators glued on the microfluidic chip *(a)*. The sensing system is a high-speed high-resolution imaging system involving an Olympus IX-71 microscope (*c*) and a high-speed camera (*letter b,* Redlake MotionXtra HG-100K, up to 100,000 frames per second). A common point of microfluidics devices involving electrokinetic pinching, segmented flow and digital microfluidics is that these techniques require actuation and detection devices orders of magnitude larger and more expensive than the chip itself, such as high-voltage power supplies, syringe pumps, drive electronics or microscopes, as seen in Figure 2. There is obviously a real need for reducing the size and cost of microfluidics actuators and sensors, exemplified by the development of simple, portable microfluidics devices in the Whitesides group [47] and by the CMOS-based sensing chips of the Shepard group [48] at Columbia University.

| Symbol | Physical property | Typical Value |
|---|---|---|
| $\gamma$ | Surface tension at the water-hexadecane interface | 52.5 mJ/m$^2$ [49] |
| $d_{31}$ | Piezoelectric strain coefficient | 190×10$^{-12}$ Pa |
| $Y$ | Piezoelectric elastic modulus | 6.2×10$^{10}$ Pa |
| $\rho$ | PZT density | 7750 kg/m$^3$ |
| $E$ | Electric field applied across actuator | 40-400 kV/m |
| $L,B,T$ | Actuator length, width, thickness | 12-20, 3-4, 0.51 mm |

Table 1. **Physical properties and typical values**

## *2. Analysis and characterization*

While our system is a first working prototype, thus far from optimal, this section provides first order analysis and characterization data that will help building next generations of devices. Section 2.1 compares the energy needed and supplied for producing the drop, section 2.2 describes the motion of the excited actuators and section 2.3 studies the relation between the pulse shape & duration and the drop generation.

## 2.1. Energy considerations

While a full study of how energy is transmitted from the moving actuator to the resulting drop via the soft rubber and the fluid in the 3-dimensional, flexible chamber is out of the scope of this article, we are comparing in this section the energy needed to form a drop with the deformation energy of the actuator. Indeed, the process of generating a microdrop of water in oil (Figure 1b) involves the sudden excitation of a piezoelectric actuator, which compresses or expands the reagent chamber. This excites the water-oil interface, which eventually breaks up and forms a drop. The minimum energy $U_d$ needed to form a drop such as the one in Figure 1b can be described as $U_d = U_s$, where $U_s = \pi d^2 \gamma$, with properties shown in Table 1, is the surface energy of the newly created drop.

The drop formation energy is provided by the flexion work of the actuator which can be approximated as $W = 0.5 F \cdot D$, where the force $F = 3d_{31}YBT^2E/(8L)$ and the maximum displacement $D = 3d_{31}L^2E/(8T)$ [50] are function of symbols described in Table 1. Also, the actuator is flexing in its first mode, with one end anchored and the other immobile along the z-direction so that $D$, the maximum z-deflection, occurs in the middle of the actuator length. For a 5 nL drop generated from a chamber with a 20mm x 3.5mm actuator, the efficiency $U_d/W$ corresponds to 0.9%, which is quite low, but is a reasonable value considering the viscous dissipation in the connecting tape, the thin PDMS layer, the chamber and the main channel fluid.

## 2.2. Motion of the actuator

Given the compliant character of PDMS, we assumed in the design process that an oscillation of relatively large amplitude would be needed to produce drops. Piezoelectric bimorph actuators have large deformations by design: for instance the out-of-plane deformation (called here the z-displacement) is on the order of tens of micrometer for an $L$=20mm long actuator [50], assuming the actuator width $w$<<L. An important parameter in the actuation design is also the eigenfrequency of the actuator, which limits the speed of deformation. The eigenfrequency $f_n$ of a piezoelectric bimorph with $L>>w$ is given in [50] for two types of boundary conditions: anchored at one end (with maximum deflection at other end) as $f_n = \frac{0.16T}{L^2}\sqrt{\frac{Y_{11}}{\rho}}$, or anchored at one end and with the other end immobile along the z-direction (with maximum z-deflection in the middle of the actuator length), as $f_n = \frac{0.48T}{L^2}\sqrt{\frac{Y_{11}}{\rho}}$. While none of these boundary conditions corresponds exactly to the experimental conditions, where one entire side of the actuator is taped on the sealing PDMS layer of the chip (see section 1), the latter was found in better agreement to the visualized motion.

Using an Optem long distance microscope objective and a high-speed camera, we measured the temporal deformation of an actuator driven by a single rectangular pulse of amplitude *dV* and duration *(2f)$^{-1}$*. Figure 3 summarizes these measurements, showing the maximum observed displacement as a function of the frequency *f* of the driving pulse. A first series of measurements, shown by empty circles, is made for a relatively large bimorph clamped at one end, with dimensions given in Figure 3. Theoretical values are also plotted as dashed lines for both the maximum static displacement and the eigenfrequency. The agreement is relatively good in terms of resonance frequency and static (low frequency) displacement. The lower resonance frequency observed experimentally can be explained by the difficulty to experimentally anchor one end of the actuator in a perfect way because we used a C-clamp. A second series of measurements is made with a smaller actuator attached via double-sided tape to a 180 μm thin PDMS layer, i.e. mounted as in the actual microfluidic chip. The two ends of the PDMS layer are then anchored firmly between two C-clamps, each clamp being about 1.5 mm away from the corresponding end of the piezo. While both the actuator size and configuration are close to the design of the microfluidic chip, the configuration is close but not exactly corresponding to the second type of boundary condition presented above: this might explain why the measured static displacement and resonance frequencies are different, both being larger than the theoretical values. Also, the recorded motion shows that the actuator ends do not move, the larger deformation occurring between these ends, confirming that the actuator vibrating in its first mode. Importantly, the visualization shows that the actuator does not freeze its motion once the driving pulse vanishes, but keeps oscillating at its natural frequency for about 6 periods, when the oscillations amplitude becomes lower than the spatial measurement error. This behavior, where the chamber experiences residual oscillations, is due to the relatively large size and inertia of the actuator, and the very soft, thin PDMS sealing layer. This behavior contrasts with existing piezoelectric drop on demand dispensers and the relative modelings [51-54], where the chamber walls are much stiffer, typically made of glass [51] or silicon [31]: in these cases the chamber deformation is in direct linear relationship to the applied voltage pulse. Obviously, any theoretical modeling of the *in-chip drop on demand* process will have to take into account the residual oscillations of the actuator and the frequency response curve of the actuator and chamber system.

## 2.3. Effects of driving pulse on the drop volumes, dispense rate & reproducibility

*In-chip drop on demand* generation is a complex fluid dynamics process involving moving solid boundaries, acoustic wave propagation and a highly deforming liquid-liquid interface. Several aspects of the drop generation process need to be optimized in order to have a good droplet generator, such as the precise control of the drop volume and motion, the elimination of smaller satellite drops, and the management of cross-talk effects in designs with multiple nozzles. Each of these aspects is sensitive to the design geometry and the actuation process [52, 54-59]. In a typical drop on demand generator, the fluidic part is made of stiff materials, which efficiently transport the pressure wave from the actuation site to the nozzle where the drop is generated. The device presented here behaves differently: the soft PDMS rubber walls reduces the apparent speed of sound in water [60] and dampens the pressure wave. This reduces cross-talk effects in chips involving multiple nozzles [61] but probably requires more energy to generate a drop (see section 2.1).

Characterization experiments reported in Figure 4 describe how the drop volume is influenced by the nozzle size, the pulse shape and the pulse duration. All the data in Figure 4 was obtained with an actuation voltage of +/- 200V. The chamber lengths used for the respective 50 and 100 µm nozzle case were 12mm and 20mm, respectively. Note that the chamber volume is about 10 nL, which is about $10^5$ bigger than a 100 pL volume drop, so that a pre-filled chamber can generate a large amount of drops of interest before refilling. After each dispense, we observed that the meniscus typically comes to the initial location within a few milliseconds. The shape of the pulse corresponds to an initial expansion of the chamber for a time $t_1$ followed by compression for a time $t_2$. Pulses with $t_1=0$ were also successful at generating a drop: they correspond to simple compression of the chamber. A quick look at the y-axis in Figure 4 shows that drops with volumes from 25 pL to 4.5 nL can be generated by varying the pulse shape and the nozzle size (corresponding to the channel height): this is a remarkable range, larger than two orders of magnitude. For a given nozzle geometry, Figure 4 also shows that the drop volume can be controlled smoothly by the pulse shape within one order of magnitude: for instance the 50 µm nozzle produces drops in the 40-300 pL range. Pulses with durations too different from an optimum duration will not produce any drop, as shown by the arrows in Figure 4. The reason might be that surface tension forces are strong enough to pull back the meniscus in the case of a short pulse or that a given pulse duration is needed to generate and amplify an unsteady pressure wave in the chamber [51, 52]. For the 50 µm nozzle we observe some dual-dispense states for pulses close to the states where no drop is ejected: a dual-dispense state correspond to a case where two smaller drops are simultaneously produced, by the *doublet instability* process described in section 3 and shown in Figure 6d. Also, in the same plot, crosses demonstrate how the drop volume can be controlled by changing the ratio between the expansion time and the compression time, while keeping the total actuation time constant. We did not systematically investigate the effect of the chamber length, but we found that a long chamber (20 mm) produces drops in an easier manner. Experiments with a middle-length chamber (12 mm) produced drops for only the maximum actuation voltage, and tests with a shorter chamber (6 mm) did not produce any drop, although the meniscus motion was clearly visible.

The results in Figure 3 and Figure 4 suggest that the optimum pulse duration to produce drops corresponds to the natural frequency $f_n$ of the actuator. Indeed, the second equation presented in section 2.2. predicts values of $f_n$ of 1.57 kHz, respectively 4.74 kHz for the actuators of the chambers with respectively the 100 µm and 50 µm nozzle. The corresponding single pulse duration for the 100 µm nozzle would be $t_2=1/(2\,f_n)$=318 µs, a time close to the 180-300 µs interval effective at producing drops in Figure 4. Similarly, the corresponding total pulse duration for the 50 µm nozzle, in the case of a pulse with $t_1=t_2$, would be $t_1+t_2=1/f_n$=211 µs, a time close to the 60-220 µs interval effective at producing drops in Figure 4. The fact that the pulse duration estimated theoretically is at the higher end of the interval of experimentally successful durations might simply indicate that the actual value of $f_n$ is slightly higher than the theoretical value, a fact shown in Figure 3 and explained in section 2.2 by the difference of boundary conditions between the experiments and the theory.

We also tried to quantify the maximum dispense rate by repeating the driving pulse with smaller and smaller time interval between pulses. Experiment shows that drops are still generated even if the time interval is reduced to 0s, which corresponds to applying the generation pulse continuously. Figure 5 and the associated movie [62] show a case where a 400 µs pulse, shaped as in Figure 4 with $t_2 = 2t_1$, is applied continuously and in turn generates a train of drops at 2.5 kHz. Note that, the nozzle is a 200 µm long and 70 µm wide straight channel. In addition, we also studied the uniformity of the drop volumes with the same nozzle. At a dispense rate of 6.2 Hz, 20 drops generated had an average volume of 1023 pL and a standard deviation of 16 pL, which corresponds to less than 2%.

## *3. Features and relevance to lab on a chip*

The process described in this paper implements in a microfluidic chip a drop on demand technique with precise, reliable control of the drop volume and generating timing. The *in-chip drop on demand* technique has the potential to perform in-chip reagent mixing, transport and multistep reactions. Four features of the *in-chip drop on demand* technique with direct relevance to lab on a chip applications are presented in Figure 6 and the associated movies, and are discussed in this section. The first feature in Figure 6a and the associated movie [63] is the ability to transport the dispensed drop away from the dispensing nozzle, along the main channel. This feature is realized by dispensing the drop in the main channel where the fluid is moved by a syringe pump: the measured drop velocity along the channel is 7 cm/s. This motion of drops by viscous drag is also used in flow focusing devices [14], and is important to the *in-chip drop on demand* technique, because transporting a droplet from the shooting area of one nozzle to another will allow multi-step reactions at frequencies of several Hertz. Interestingly, Figure 6a shows that a smaller particle is embedded in the main drop, a phenomenon that can be suppressed or encouraged by adjusting the actuation pulse shape and intensity. Also, the ability of dispensing a drop that encapsulates another particle could be of interest for manufacturing complex multi-wall or hollow spheres, a topic of recent interest in microfluidics [64, 65]. Similar considerations can be made for the smaller satellite drop generated between the drop and the nozzle. Issues related to satellite formation have received abundant attention [59, 66] because of their relevance to printing quality.

Figure 6b and the associated movie [67] show the second feature, which is the ability to digitally control the dispensed drop volume, by generating additional drops that coalesce with the original drop. The first frame shows a 500 pL drop, the volume of which increases to 3.5 nL by 6 successive increments of 500 pL. This coarse, digital way to control the drop volume can be coupled with the finer, analog volume control of modifying the pulse parameters (section 2.3), in order to exactly dispense the desired quantities over a wide range of volumes. While in-flight coalescence has been realized [68, 69] by *atmospheric* drop on demand, where drops are jetted in the air, the *in-chip drop on demand* technique allows a simpler realization of coalescence because the dispensed drop becomes immobile in the main channel after the kinetic energy of the dispense has been dissipated. The attentive reader probably noticed that the channel walls in Figure 6(a-b) are irregular: indeed these nozzles were manufactured during preliminary experiments where the master was simply a piece of electric tape applied to a glass slide and approximately cut to the desired geometry with a sharp cutter under the stereo microscope. Finally, it is worth mentioning that the large drops created in Figure 6b can be split into smaller drops by moving them into a T-shaped connection [47].

The potential of mixing different reagents into a single drop is illustrated in Figure 6c and the associated movie [70]. The nozzle on the left generates a drop of ink, while a pure water drop is generated by the right nozzle simultaneously and hits the ink drop. Coalescence occurs then at *t*=22ms, starting the mixing of the ink and water through diffusion and the transient flow associated with coalescence and the drop transport along the channel. Interestingly, coalescence does not occur right after the drops hit each other, probably because of the thinning and breakup of an oil film between the two drops. This delay could be reduced by oppositely charging the two drops [23]. It is also worth mentioning that mixing and particle transport can be controlled by vibrating the liquid [71] with the same piezoelectric actuators that generate the drops. To some extent, the mixing process presented here can be compared to the impressive *airborne chemistry* technique, where a drop is immobilized in the air at the node of a high-power ultrasound field: this main drop acts then as an isolated reactor fed by smaller drops of reagent dispensed by atmospheric drop on demand nozzles. Airborne chemistry has been successful for screening the conditions for protein crystallization or for performing biological analyses [72-74]. Similarly, *in-chip drop on demand* also allows the drop dispensed in an immiscible fluid to function as an isolated reactor, fed by further reagent additions from neighboring nozzles. Major differences with the airborne chemistry technique are that optical measurements might be more difficult with the present technique, due to the presence of the hexadecane and the PDMS wall; however the surrounding hexadecane allows higher heat transfer and suppresses evaporation.

Figure 6d and the associated movie [75] show the fourth feature, which is the ability to generate a *doublet* of drops, while applying a single excitation pulse to the actuator. This occurs when an initially generated drop is hit by a strong subsequent excursion of the meniscus. During the process, the meniscus breaks the initial drop into two half drops while briefly assuming the shape of a well-known cartoon character (367 µs). We call this type of dispense the *doublet* dispense. To the best of our knowledge, no *doublet* dispense has ever been realized with atmospheric drop on demand techniques because the dispensed drop quickly travels away from the nozzle area where the meniscus oscillates.

Finally, an interesting feature of the proposed *in-chip drop on demand* technique is that the frequencies needed to produce drops are on the order of a few kHz (see section 2.3), right into the audio frequency domain. Therefore, one should be able to replace the research-grade pulse generator and amplifier used to drive the actuators by inexpensive audio components: for instance, audio amplifiers are mass-produced and offer multi-channel capabilities, up to eight channels for a $300 home cinema amplifier. We tested this hypothesis by powering the microfluidic chip with a used audio home stereo

amplifier (JVC AX-R87, 4 channels, 400W, $37 on a popular auction site), shown with the letter *g* Figure 2. The pulses amplified by the audio amplifier showed minimal noise and a similar shape as the pulses amplified by the research-grade Krohn-Hite amplifier, and we managed to produce a single drop on demand (movie [76]) by driving the actuator with the home audio amplifier connected to the function generator (d) in Figure 2. This ability to drive a microfluidic chip with inexpensive, mass-produced audio electronics might seem anecdotal, however it is our conviction that the microfluidics growth will benefit from low-cost actuation schemes, which strongly contrast with the expensive components used at present.

In addition to dispensing drops, the piezoelectric actuation technique described here can be used to dispense a single bubble on demand in the main channel as shown in Figure 7 and the associated movie [77]. The bottom chamber and horizontal channel are filled with water, while the top channel is filled with air. Piezoelectric actuation of the bottom channel is used to modify the volume of the water chamber and pull the air out, which breaks up into a single bubble with the help of finger-like features next to the nozzle.

## *Conclusion and Outlook*

The *in-chip drop on demand* technique presented in this article allows individual dispensing of drops of aqueous reagents or bubbles in a microfluidic chip with a temporal precision of one millisecond, and very high rates. The ability to precisely trigger the drop generation time will allow the coordination of the generation of drops with events occurring in the chip, such as the detection of chemical reaction or temperature changes, or the transit of biological cells and other particles. The drop volume can be controlled from 40 pL to 4.5 nL by varying the pulse shape, the chip geometry, or by merging several drops together. The generated drop is surrounded by an immiscible fluid, which prevents evaporation, enhances heat transfer and can be used to transport the drop by viscous drag. Interestingly also, the dead volume is quite small: a typical chamber filled with a few microliter will dispense several thousands drops with a typical volume of 100 picoliter.

In terms of flexibility and individuality, the proposed *in-chip drop on demand* technique is comparable to the digital microfluidics technique, with the advantages to work with any aqueous fluid and not only dielectric fluids, and to dispense smaller drops. In terms of mixing speed and ability to encapsulate reagents within an immiscible liquid, the *in-chip drop on demand* technique is comparable to segmented flow techniques, while offering more flexibility because each single droplet generation event is triggered and controlled. Finally the ability to drive the system with inexpensive, mass-produced audio electronics is demonstrated, a feature that might help the commercial adoption of this technology.

## *Acknowledgements:*

This work has been supported partially by NSF grants 0449269 and 0701729. The last author warmly thanks his academic father, Dimos Poulikakos, who introduced him to the joy of academic research and specifically to the drop on demand technique, together with David Wallace from Microfab.

**Figures**

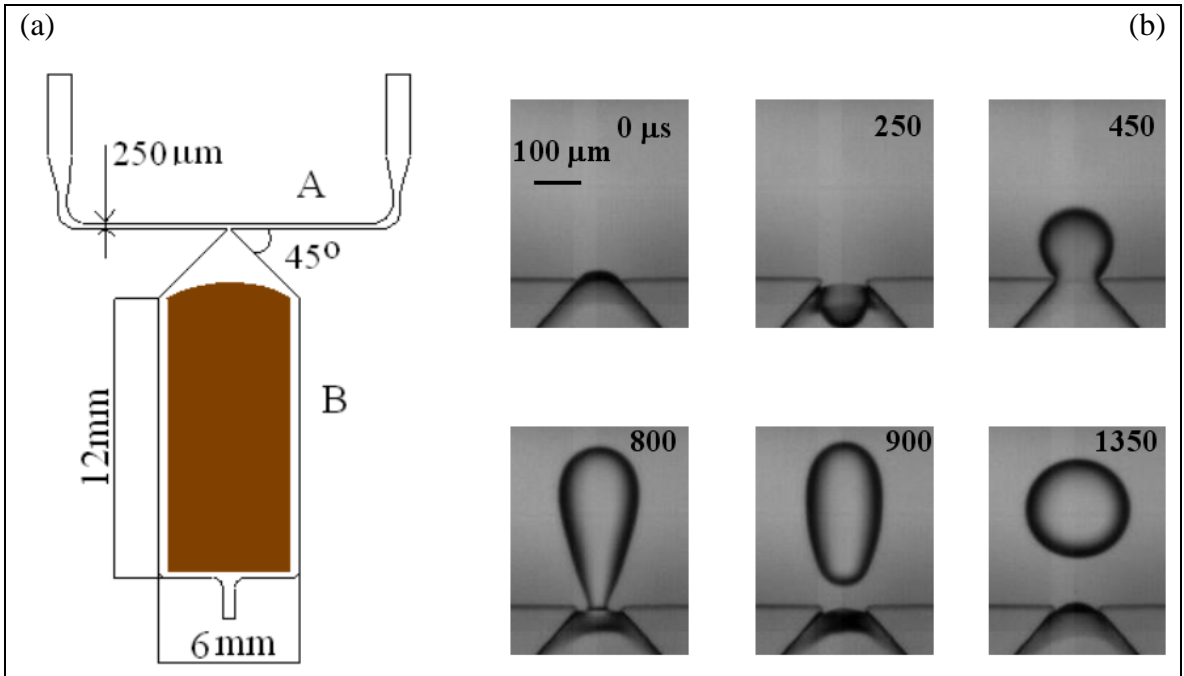

Figure 1: (a) Geometry of a microfluidic chip for *in-chip drop on demand* dispensing. The bottom chamber is filled with an aqueous reagent. A piezoelectric bimorph actuator glued to the chamber allows the release of an aqueous drop on demand in the horizontal channel filled with an immiscible fluid. (b) Stages depicting the formation of a 1nL drop from a 50 μm nozzle, also shown in the associated movie [44].

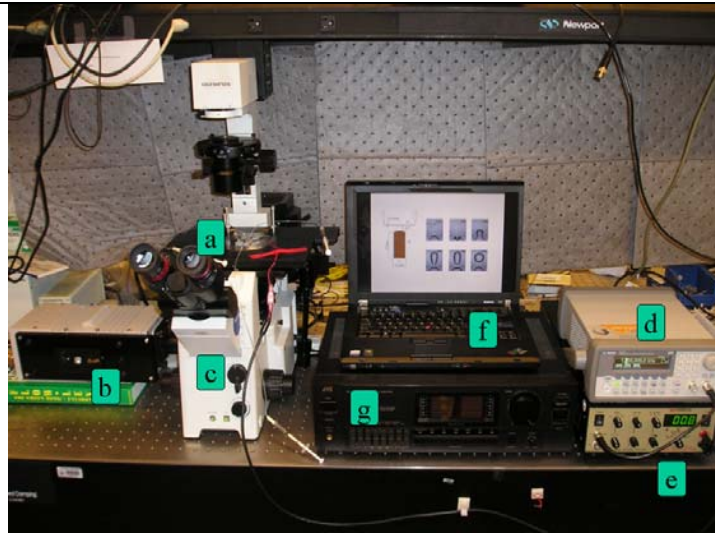

Figure 2: Description of the experimental setup. The microfluidic chip (a) is actuated either by a function generator (d) and a high-voltage amplifier (e), or by the soundcard of computer (f) and an audio amplifier (g). Sensing is performed by the high-speed camera (b) and the inverted microscope (c). Note the scale paradox between the tiny microfluidic chip and the bulky actuation and sensing components, typical of current microfluidic systems.

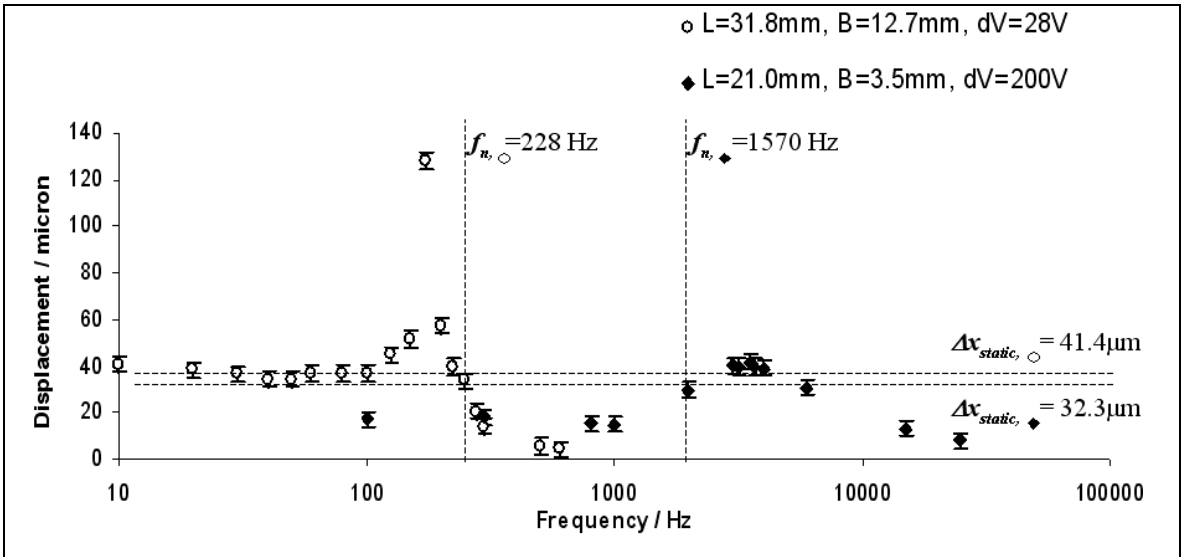

Figure 3: Influence of the excitation frequency on the amplitude of the actuator motion. The empty circles and full lozenges denote two types of boundary conditions as described in section 2.2. The actuator size and pulse amplitude are given in the legend. The dashed lines denote the theoretical values for natural frequency and static displacement.

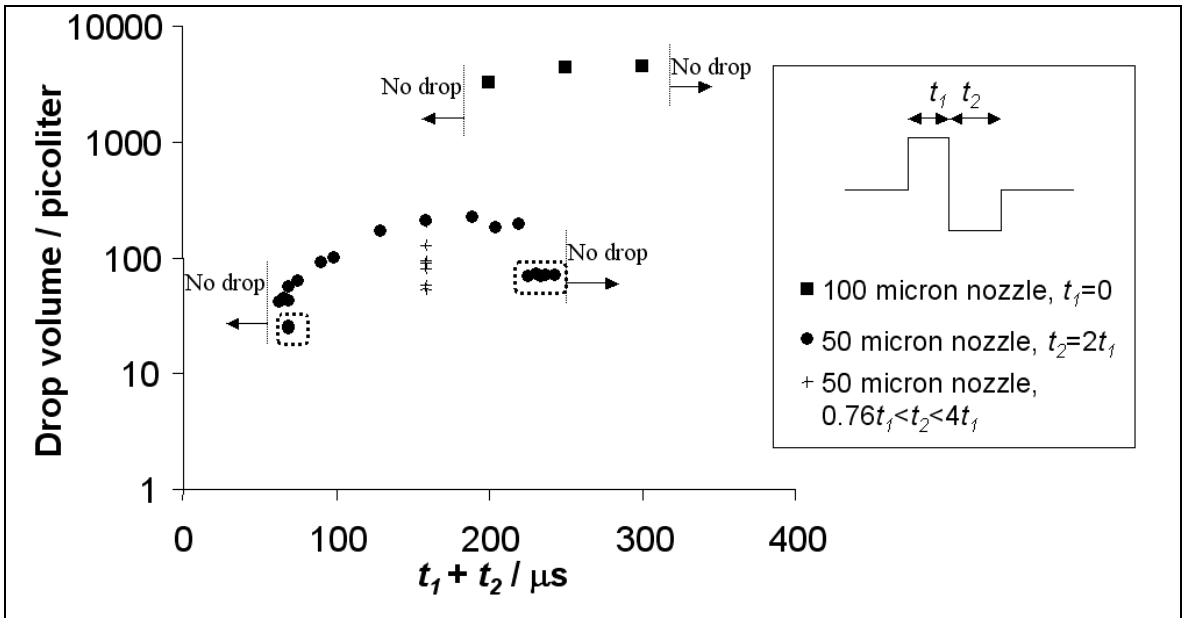

Figure 4: Drop volumes as a function of dispense parameters (nozzle size, pulse shape or length). The horizontal axis denotes the total pulse length, which involves the chamber expansion followed by the chamber compression, with respective duration $t_1$ and $t_2$. The nozzle width is indicated in the legend, with channel height equal to the same as the nozzle width. The dotted zones show *doublet* dispenses, where two drops of smaller volume are generated simultaneously by a single pulse [75].

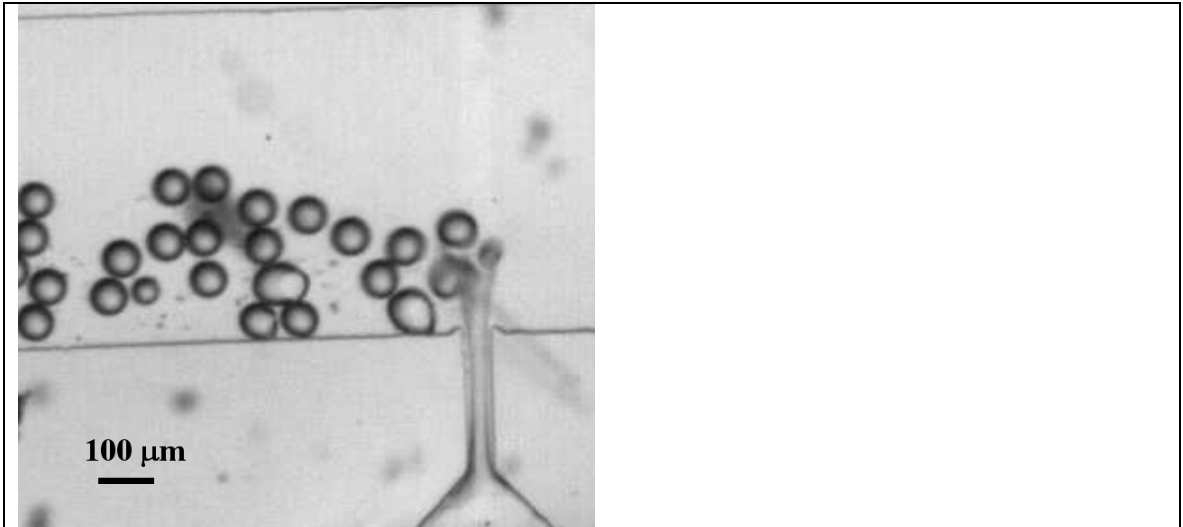
Figure 5: Dispense of individual drops by repeating the pulse at a rate of 2.5 kHz [62].

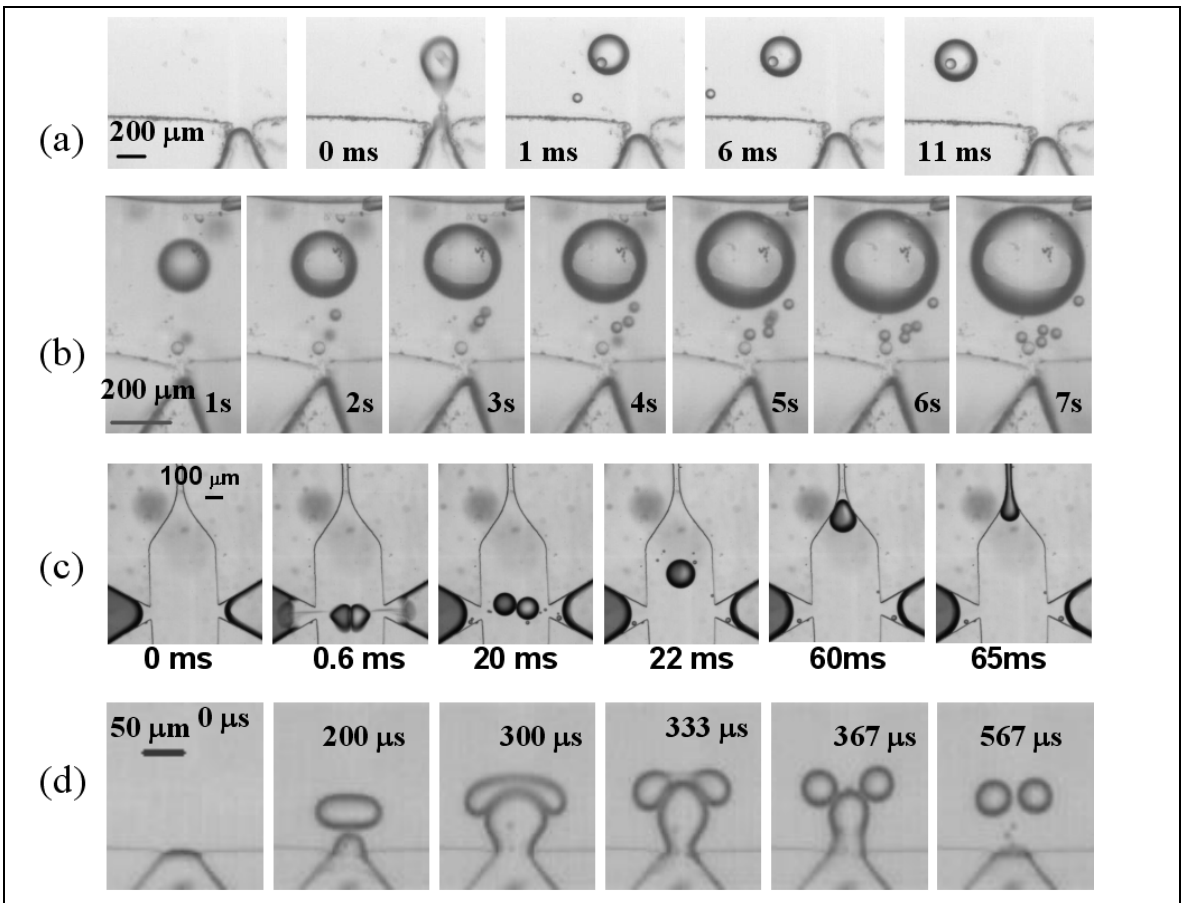
Figure 6: Four features of *in-chip drop on demand* with relevance to lab on a chip applications. The frames are extracted from the associated movies. (a) drop transport by viscous drag [63], (b) digital control of drop volume [67], (c) merging and mixing of two different reagents [70], and (d), *doublet* dispense, where two drops of small volume are generated simultaneously by a single pulse [75].

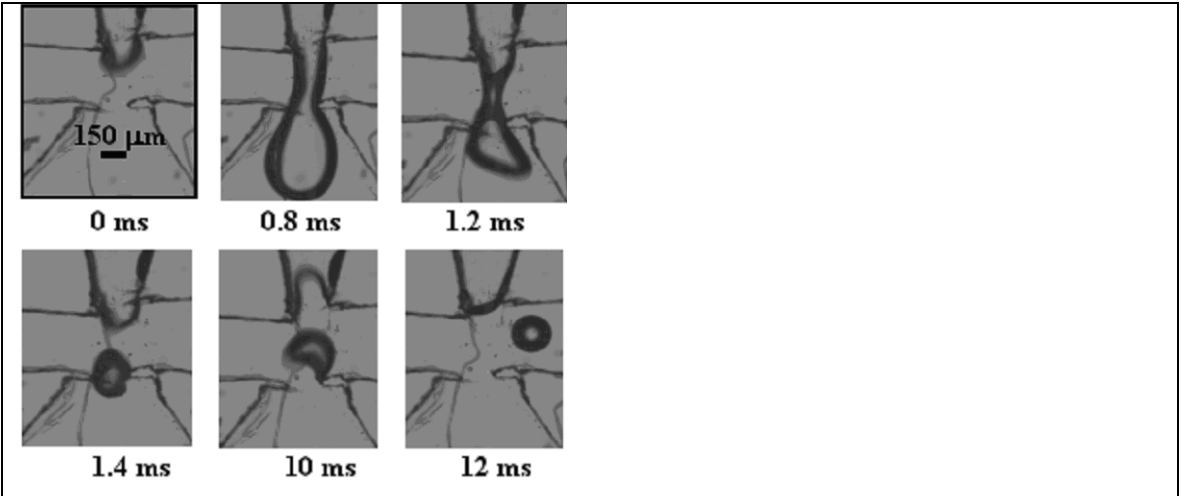

Figure 7: A 1nL gas bubble is generated on demand in a microfluidic chip [77]. There is no flow in the main channel.